# Non-adiabatic Molecular Dynamics Simulation for Carrier Transport in a Molecular Monolayer

*Junfeng Ren*[(1,2)], *Nenad Vukmirović*[(3)], *Lin-Wang Wang*[(1)*]

(1) Material Science Division, Lawrence Berkeley National Laboratory, One Cyclotron Road, Mail Stop 66, Berkeley, CA 94720

(2) College of Physics and Electronics, Shandong Normal University, Jinan 250014, China

(3) Scientific Computing Laboratory, Institute of Physics Belgrade, University of Belgrade, Pregrevica 118, 11080 Belgrade, Serbia

(*) Email: lwwang@lbl.gov

**ABSTRACT**: We present a new approach to carry out non-adiabatic molecular dynamics to study the carrier mobility in an organic monolayer. This approach allows the calculation of a 4802 atom system for 825 fs in about three hours using 51,744 computer cores while maintaining a plane wave pseudopotential density functional theory level accuracy for the Hamiltonian. Our simulation on a pentathiophene butyric acid monolayer reveals a previously unknown new mechanism for the carrier transport in such systems: the hole wave functions are localized by thermo fluctuation induced disorder, while its transport is via charge transfer during state energy crossing. The simulation also shows that the system is not in thermo dynamic equilibrium in terms of adiabatic state populations according to Boltzmann distribution. Our simulation is achieved by introducing a linear time dependence approximation of the Hamiltonian within a fs time interval, and by using the charge patching method to yield the Hamiltonian, and overlapping fragment method to diagonalize the Hamiltonian matrix.

MANUSCRIPT TEXT

1. Introduction

Carrier transport in an organic system is a complex phenomenon which can involve different underlying mechanisms [1,2]. These include: delocalized bulk band structure transport [3,4], localized state single phonon absorption/emission hopping [5], atomic relaxation induced localized polaron state hopping [6,7], polaron band structure [8], thermo fluctuation induced dynamic disorder transport [9,10], and coherent electron transport (which also includes the band structure transport and tunneling transport) [11,12]. These different mechanisms give rise to rich features in terms of the temperature, carrier density and electric field dependences of the carrier mobility [2,13,14]. Most of the above underlying mechanisms involve electron-phonon coupling, some strongly coupled, some weakly coupled, some related to the diagonal (same electron state) coupling [6], some off diagonal coupling [5]. Traditionally, different models with different approximations are used to describe different carrier transport phenomena. Unfortunately, it is often not known a priori which mechanism is involved for a given system. Furthermore, it is possible that several mechanisms are all at play in the same time, or they dominate at different temperature regimes for the same system. Thus, it will be helpful to treat all these mechanisms under an uniform framework. One such approach is the non-adiabatic molecular dynamics (NAMD) [15,16,17].

In the NAMD, the electron and the nuclei are moving at the same time. While the nuclei are moving following their classical Newton's law, the electron wave functions are evolving following the time dependent Schrodinger's equation. By moving both electron and nuclei, and including their interaction through the Hamiltonian, all the aforementioned mechanisms can be described by NAMD. In the conventional ab initio molecular dynamics (ABMD) [18], at any given time t, the electrons will occupy the electronic ground state for the atomic configuration at that time (the Born-Oppenheimer approximation). In NAMD, this is no longer true. The electronic state at time t will depend on its history, and it is described by the time dependent Schrodinger's equation.

The applicable scope of NAMD is far beyond the carrier dynamics in organic systems. For example, it can be used to study the carrier cooling from high excited states [19], charge transfer between organic and inorganic systems [20,21], and coherent or semicoherent processes [11]. It can also be used to describe the effects of charge transfer to the ionic movements in a catalytic process [22], or charge transfer in a solvent [23]. It has also been used to study the dynamic behavior in multi-electron systems [24].

In recent years, we saw a surge of NAMD studies [25,26,27,28,29], from organic systems to nanostructures, showing the promise and people's interest of this approach. Unfortunately, the NAMD simulation can be extremely expensive. Due to the small mass of the electron (equivalently, the large energy spectrum of its movement), the time step one can use to evolve the Schrodinger's equation is typically $10^{-3}$ fs. This is a thousand times shorter than the traditional ABMD step. As a result, the current NAMD calculations are either for relatively small systems (e.g., less than 200 atoms) [19,20,21], or for tight-binding [25] and other simplified model Hamiltonians [30,31].

In the current paper, we present a few new techniques which significantly accelerate the speed of the NAMD simulation. Using these techniques, for the first time, a NAMD simulation has been carried out for a 4802 atom system for about 1 ps using an ab initio Hamiltonian. The Hamiltonian is described using the plane wave pseudopotential (PWP) method at the density functional theory (DFT) level. We have used this simulation to study the carrier transport mechanism for a monolayer of five thiophene ring oligomer 5TBA (pentathiophene butyric acid), which demonstrates a herringbone 2D crystal structure. Our simulation reveals a previously unknown new mechanism for carrier transport in the system: a thermo fluctuation induced wave function localization, and the charge transfer during state energy crossing. We expect similar mechanism can be applied to some other organic crystals.

The highlights of our study can be summarized as the following: (i) a linear change approximation for the electron Hamiltonian is used within a time interval of 0.5 fs. This allows the time step for the computationally most expensive part to be 0.5 fs, instead of $10^{-3}$ fs, thus increases the speed of the simulation by hundreds of times. (ii) a modified Ehrenfest dynamics is introduced, which can be used as an alternative method to Tully's energy surface hopping algorithm. (iii) in the current work, we have ignored the feedback from the electron movement to the nuclei movement, thus the nuclei movement is detached from the electronic state description and can be described by a classical force field. This approximation is often used to treat the carrier dynamic in a large system and when the polaronic effect is relatively small. (iv) the charge patching method (CPM) is used to describe the Hamiltonian, while an overlapping fragment method (OFM) is used to diagonalize the Hamiltonian matrix. Although these are approximations, we show that the errors they introduced are rather small. (v) we found that the main mechanism for the carrier transport in the 5TBA thin film is by the state crossing, while the state localizations are induced by thermo fluctuation (dynamic disorder), rather than polaronic atomic relaxation. This is a new mechanism which has not been discussed before. It is different from the polaronic state crossing (Marcus theory [32]), and also different in some degree from the pure dynamic disorder picture proposed by Troisi et.al. [9].

2. **The formalism and the approximations**

Here we describe the basic formalism for a NAMD simulation. If we use {R(t)} to denote the nuclei positions, which are changing as functions of time t, then the electron Hamiltonian H, which depends on R(t), is also a function of time. As a result, we have a time dependent Schrodinger's equation (the Kohn-Sham equation in DFT) as:

$$i\frac{\partial \psi}{\partial t} = H(t)\psi \quad (1)$$

Here we have used ψ to denote a single electron wave function. Meanwhile, the molecular dynamics (MD) movement of {R(t)} is described by the classical Newton's equation:

$$M_i \frac{d^2 R_i(t)}{dt^2} = F_i \quad (2)$$

Here $M_i$ is the mass of atom i, $R_i(t)$ is its position and $F_i$ is the force acting on atom i. The force can be calculated ab initio from DFT method based on H(t). In that case, the forces depend on the single electron wave functions ψ, thus equations (1) and (2) are fully coupled. As we mentioned in the introduction, in the current work, we will use classical force field (CFF) method to describe the nuclei dynamic of the organic system. As a result, $F_i$ only depends on the atomic configuration {R(t)}, thus while Eq.(1) depends on Eq.(2), Eq.(2) no longer depends on Eq.(1). This detachment has been used, for example, to study the carrier cooling in a quantum dot [19]. There are some consequences of this approximation. First, the total energy of the electron+nuclei system is no longer conserved. For example, the nuclei system can continuously pump energy into the electron system, while the energy of the nuclei system itself remains conserved. However, this problem can be partially solved by the Tully's algorithm with some modifications. In this paper, we also introduce an alternative method (to be called modified Ehrenfest dynamics) which introduces a Boltzmann factor in the state-to-state transition rate. Although we still cannot define a conserved total energy after these treatments, they will prevent the system from divergence in total energy, and help it to reach some quasi steady state. That is sufficient for the purpose of charge transfer calculations. The second problem is that the detachment will ignore the polaronic atomic relaxation. In this effect, the occupation of one localized electron state will induce atomic relaxations near that state, which further localizes the electron wave function. We have tested that in our current system, such polaronic effect is relatively small. Nevertheless, for general problems, we do plan to include such polaronic effects in the future work with some approximated methods.

To solve Eq.(1), one can just apply H directly to wave function ψ at every time step. But that can be quite time consuming. Furthermore, as we will show later, it is necessary to introduce energy surface hopping, or other state transition analysis. This requires us to analyze the wave function in terms of the adiabatic states {$\varphi_i(t)$}, which are eigen states of H(t) at any given time t:

$$H(t)\varphi_i(t) = \varepsilon_i(t)\varphi_i(t) \tag{3}$$

Thus a common practice is to expand the wave function ψ(t) using these adiabatic states:

$$\psi(t) = \sum_i C_i(t)\varphi_i(t) \tag{4}$$

Then the solution of Eq.(1) becomes the solution of coefficients {$C_i(t)$}. One can plug Eq.(4) back to Eq.(1), then we have:

$$\dot{C}_i(t) = -i\varepsilon_i(t)C_i(t) - \sum_k C_k(t)V_{ik}(t) \tag{5}$$

and here $V_{ik}$ is a transition rate between adiabatic states "i" and "k", which can be calculated as:

$$V_{ik}(t) = \left[\langle \varphi_i(t) | \varphi_k(t+\delta t) \rangle - \delta_{i,k}\right]/\delta t \tag{6}$$

The Hamiltonian in Eq.(1) depends on atomic positions {R(t)}, but it also depend on the electron wave functions ψ. In the current study, we will only consider the situation where only one carrier (in our case, a hole in the valence band) exists, which is not in its ground state (e.g., the valence band maximum (VBM) state for the hole), and all the other valence electrons are in their ground state. Thus, the total charge density of the system at t equals the ground state charge density $\rho_0(t)$ of the neutral system plus (or minus in our case of hole) $|\psi(t)|^2$. However, if the carrier wave function ψ(t) is not very localized, for the same reason that we have ignored the effect of ψ(t) to the nuclei force F, here we can also ignore the contribution of ψ(t) to the Hamiltonian H(t). Note, as we are using H to describe other electron's interaction to the carrier wave function ψ in Eq.(1), ignoring the effect of $|\psi(t)|^2$ to the Hamiltonian can also help us to avoid the self-interaction error [33] in the local density approximation (LDA).

Thus, we will use the ground state (Born-Oppenheimer energy surface) charge density to describe H, which then depends only on {R(t)}. One can occupy all the valence adiabatic states {$\varphi_i(t)$} from Eq.(3) and solve the Hamiltonian H self consistently as in a traditional ABMD to get H(t). Here we will use the charge patching method (CPM) [34,34] to get $\rho_0(t)$ for a given atomic configuration {R(t)}. The CPM [34,35] is a well tested method to provide the ground state electronic charge density without going through a self consistent calculation. It generates atomic charge density motifs from small system calculations, then patches these motifs together to get the charge density of a large system. After $\rho_0(t)$ is obtained, it is used with the DFT formalism (e.g., LDA) to get the single particle Hamiltonian (the total potential V(r)). The resulting eigen energy error of a CPM is typically 20-30 meV compared with the direct LDA calculation [35]. Figure 1 shows the comparison between the CPM and the direct self consistent field (SCF) LDA ground state result for one snapshot {R(t)} of a 2x2 5TBA monolayer supercell (each unit cell has 2 5TBA oligomers, thus there are 392 atoms in the supercell) with room temperature random thermo atomic displacements. Table.I also lists the first few hole eigen energies of a few snapshots for the 2x2 system comparing the SCF LDA results with the CPM results. From both Fig.1 and Table.I, we see that the CPM eigen energies are close to the SCF LDA eigen energies. This establishes the CPM as a good method to describe the ground states of our system.

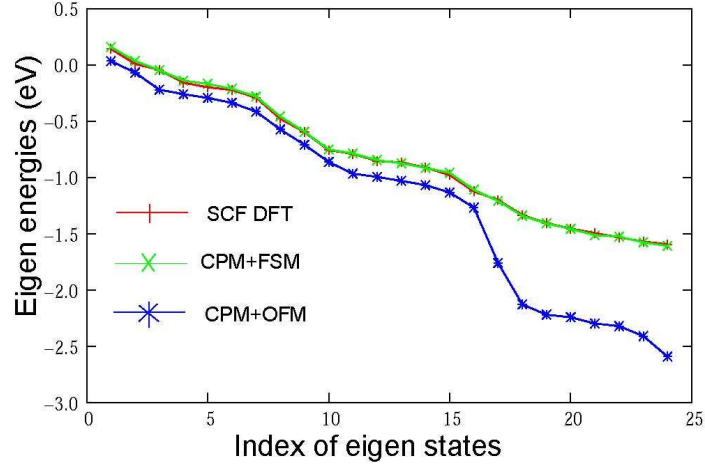

Fig.1, comparison of the eigen energies between SCF DFT, the CPM, and the CPM solved with OFM. Note the folded spectrum method (FSM) is an exact method involves no approximation for solving the electron eigen states for a given H.

|  | Method | VBM (eV) | VBM-1 (eV) | VBM-2 (eV) |
|---|---|---|---|---|
| T=0 K Relaxed | SCF DFT | 0.095 | -0.066 | -0.148 |
|  | CPM | 0.092 | -0.068 | -0.150 |
| T=300K Snapshot 1 | SCF DFT | 0.153 | 0.008 | -0.056 |
|  | CPM | 0.174 | 0.023 | -0.008 |
| T=300K Snapshot 2 | SCF DFT | 0.143 | 0.008 | -0.046 |
|  | CPM | 0.158 | 0.034 | -0.047 |

Table I, the eigen energies comparison between the selfconsistent DFT calculations (SCF DFT) and the charge patching method calculations (CPM) for different atomic configurations of a 2x2 herringbone structure supercell.

Using CPM, at any given time t, with {R(t)} provided by the CFF MD simulation, we can construct the Hamiltonian H(t) relatively easily. Now, the task is to integrate Eq.(5). To do that, one needs a small time step of $10^{-3}$ fs. This requires the evaluation of $\varepsilon_i$ and $V_{ik}$ at every $10^{-3}$ fs interval. If they are calculated by directly solving the Schrodinger's equation (3) for every $10^{-3}$ fs, it will be prohibitively expensive in computation. Here, we have adopted a linear approximation for H(t). According to this approximation, within a time interval [$t_1$, $t_2$] (here $\Delta t = t_2 - t_1$ is of the order of fs), the Hamiltonian H(t) will have the following linear dependence on time t:

$$H(t) = H(t_1) + (t - t_1)(H(t_2) - H(t_1))/(t_2 - t_1) \qquad (7)$$

To test this approximation, we have represented H(t) in the basis of the adiabatic eigen states at $t_1$: $\{\varphi_i(t_1)\}$. Some of the typical matrix elements: $\langle \varphi_i(t_1) | H(t) | \varphi_j(t_1) \rangle$ are shown in Fig.2. As we can see, these matrix elements are approximately linear within a 0.5 fs time interval. Note, although the Hamiltonian H(t) changes linearly within this interval, the adiabatic eigen states $\varphi_i(t)$ do not, for example due to rapid state energy crossing. Now, we first solve Eq.(3) for every $\Delta t$ (0.5 fs). From that, we have $\{\varepsilon_i(t_1), \varphi_i(t_1)\}$ and $\{\varepsilon_i(t_2), \varphi_i(t_2)\}$. In our cases, we have solved 50 top valence band states near the band edge for the 7x7 supercell system. These 50 states span an eigen energy window of about 0.35 eV. We expect that this set of adiabatic eigen state is large enough for Eq.(5); as we will show later, the average energy of $\psi(t)$ is typically only about 0.05 eV away from the VBM energy, and the coefficient $C_{50}(t)$ in Eq.(5) is extremely small (in the order of $10^{-11}$ to $10^{-15}$).

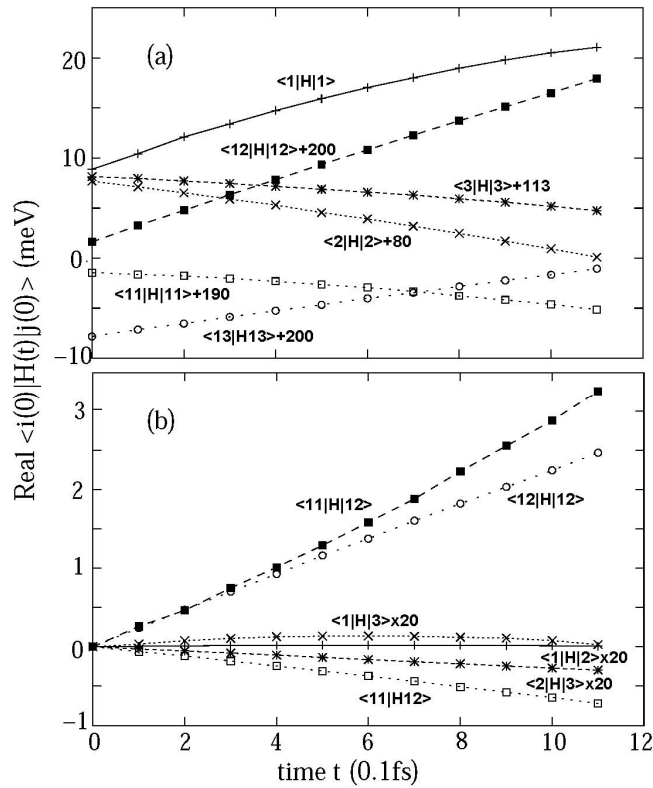

Fig.2, the matrix elements of H(t) under the basis set of $\{\varphi_i(t_1)\}$. Here the number in the Dirac bracket is the index "i" of the basis function. The +200, +113, +80, +190 in (a) are the amount of shifts in meV in order to bring these curves together for viewing. This is for the 7x7 supercell system.

Now, we will use $\varphi_i(t_1)$ i=1,N (N=50) as the basis set to diagonalize H(t). To do that, we need the matrix elements $\langle\varphi_i(t_1)|H(t)|\varphi_j(t_1)\rangle$. Obviously we have $\langle\varphi_i(t_1)|H(t_1)|\varphi_j(t_1)\rangle = \varepsilon_i(t_1)\delta_{i,j}$. Therefore, all we need is to get the matrix element $\langle\varphi_i(t_1)|H(t_2)|\varphi_j(t_1)\rangle$ in order to use Eq.(7) to get $\langle\varphi_i(t_1)|H(t)|\varphi_j(t_1)\rangle$. However, we know $\langle\varphi_i(t_2)|H(t_2)|\varphi_j(t_2)\rangle = \varepsilon_i(t_2)\delta_{i,j}$ Let's now assume {$\varphi_i(t_1)$} can be expanded by {$\varphi_i(t_2)$} as: $\varphi_i(t_1) = \sum_{j=1,N}\langle\varphi_j(t_2)|\varphi_i(t_1)\rangle\varphi_j(t_2)$, then $\langle\varphi_i(t_1)|H(t_2)|\varphi_j(t_1)\rangle$ can be obtained from $\langle\varphi_i(t_2)|H(t_2)|\varphi_j(t_2)\rangle = \varepsilon_i(t_2)\delta_{i,j}$ by a simple unitary transformation. In reality, the transformation $\varphi_i(t_1) = \sum_{j=1,N}\langle\varphi_j(t_2)|\varphi_i(t_1)\rangle\varphi_j(t_2)$ might not be unitary due to the lack of completeness of the basis set. We have carried out a Gram-Schmidt orthonormalization to make the matrix $\langle\varphi_j(t_2)|\varphi_i(t_1)\rangle$ unitary. However, most of the modifications happen to high eigen states i (when i is close to N). Because their coefficients $C_i(t)$ in Eq.(4) are very small, the overall effect of this procedure on $\psi(t)$ is very small.

Now, we can integrate Eq.(5) within the interval [$t_1,t_2$]. This time, a very small time step dt (~$10^{-3}$ fs) can be used. At every small time step dt, the small NxN matrix $\langle\varphi_i(t_1)|H(t)|\varphi_j(t_1)\rangle$ is diagonalized. This gives us {$\varepsilon_i(t),\varphi_i(t)$}, which can be used to evaluate Eq.(6) and integrate Eq.(5). It is interesting to comment that, we believe the similar linear approximation on H can be used for real time simulations of time-dependent DFT (TDDFT) [36]. There, the time dependent wave functions for all the occupied states need to be calculated. Nevertheless, depending on the situations, the related N can still be much smaller than the dimension of the original Hamiltonian, thus significant saving becomes possible.

By using the linear approximation of H(t) in Eq.(7), we have increased the time step for solving the expensive Schrodinger's equation (3) from dt ($10^{-3}$ fs) to $\Delta t$ (0.5 fs). Thus, the new computational cost is similar to that of a conventional ABMD. Furthermore, we have used the CPM to construct H(t), therefore avoiding the need to do SCF DFT calculations at every step of $\Delta t$. This allows us to carry out the NAMD simulation for a 4802 atom system (a 7x7 5TBA thin film supercell) for a long time (~1 ps). We can solve Eq.(3) using the folded spectrum method (FSM) [37], which is exact, and does not introduce any additional approximation. However, we found that it is faster to use the overlapping fragment method (OFM) [38] to diagonalize Eq.(3) especially because we need to solve N=50 eigen states. The OFM method has been used to solve the eigen states of organic polymers, and yielded good results compared with the direct FSM calculations [38]. In our case, in the OFM calculation, each 5 ring 5TBA oligomer has been cut into three mutually overlapping fragments. Each fragment contains three thiophene rings, with the cut off bond passivated by additional H atoms. The eigen states of each fragment are solved separately (with its charge density calculated by the CPM). Then, the top of the valence band state of each fragment is used as the basis set to diagonalize the original full system

Hamiltonian H(t). The detail procedure of the OFM is described in Ref.[38]. In Fig.1, we also show the OFM results for the 2x2 supercell system. We can see that, although there is a 150 meV overall shift for the $\varepsilon_i$ curve, the overall shape of this curve agrees quite well with the original CPM+FSM result, and the SCF DFT result. Large errors only occur when the eigen energy is 1.5 eV away from the band edge (where the adiabatic eigen states are no longer included in the $\varphi_i(t)$ of Eq.(4). The N=50 eigen states $\varphi_i(t)$ in a 7x7 system only span a 0.35 eV energy window). We have also tested the 7x7 supercell system used for our final study. There are 294 fragment basis functions in the OFM. The OFM and FSM solved eigen states look similar for their spatial localizations. The eigen energy changes with time are also similar for these two methods. For example, we have taken two snapshots in the MD trajectory, the VBM eigen energy difference between these two snapshots is -0.12 eV using FSM, while it is -0.14 eV using OFM. Thus, we believe the OFM can quantitatively describe the eigen states of the system, and can be used for our NAMD simulation.

### 3. The physical systems, surface hopping and modified Ehrenfest procedures

We have used the formalism described in section 2 to calculate a 7x7 supercell of a 5TBA monolayer thin film. The structures of the thin film at room temperature and the single 5TBA oligomer are shown in Fig.3. This is a system similar to a previous tetradecyl pentathiophene butyric acid (TD5TBA) monolayer [39] system studied by atomic force microscopy. The 5TBA monolayer has also been made experimentally, and its p-type carrier lateral mobility is planned to be measured using micro-electrodes [40]. Both TD5TBA and 5TBA monolayer exhibit a herringbone packing pattern viewed from the top as shown in Fig.3. 5TBA oligomer differs from TD5TBA by missing the alkane chain on the top of the molecule. As a result, the 5TBA oligomers tend to stand up vertically within the monolayer, instead of leaning towards one side as in the TD5TBA case [39]. The force field calculated 5TBA lattice constant in the x and y directions (Fig.3) are 7.6 and 5.7 Å respectively, which agree well with the experiments [40].

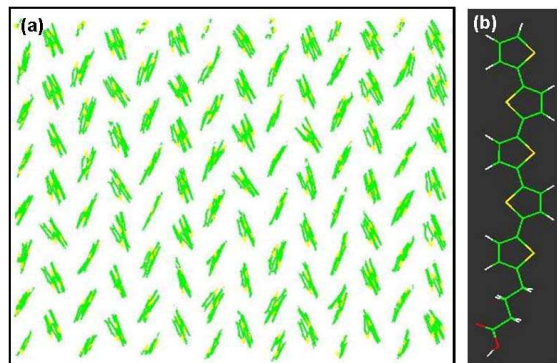

Fig.3, top view of the herringbone pattern (a) of the 5TBA oligomer (b) (side view). A 7x7 supercell with 4802 atoms is used in our study. In (b), the yellow indicates the S atom, green the C atom, white the H atom and red the O atom. The classical

force field is used in a MD simulation to provide the thermo fluctuation of the molecules at room temperature. The horizontal direction is defined as the x direction while the vertical direction is defined as the y direction. The dimensions for these two directions for the 7x7 supercell are 53.20, 39.90 Å respectively.

The system in Fig.3 is simulated using CFF91 [41] force fields with some parameters modified to make the relaxed 5TBA molecule structure agrees with the DFT result. Verlet algorithm is used to carry out the molecular dynamics using the code LAMMPS [42]. The simulation is done for several ps at the room temperature [T=300K], while the last ps is used in our NAMD calculation.

Using the R(t) obtained from the classical MD, we then performed the NAMD calculation according to Eqs.(3)-(7) and the procedure described in section 2. Fig.4 shows the first few adiabatic eigen state energies changing with time t, and the isosurface plots of the corresponding eigen states. From Fig.4, we can see that there are a lot of state energy crossings, especially for the lower energy valence states. Sometime, one can trace one eigen state, seeing that its location does not change much although its energy can go up and down over 0.1 eV (e.g., the second valence band (VB) state at t=0, the pink state, which is the same state as the third VB state at t=12 fs, the golden state. Here, when we trace on "one state", we are not referring to one "i" in $\varphi_i$, instead, we are connecting the states after a state crossing by their wave function similarity).

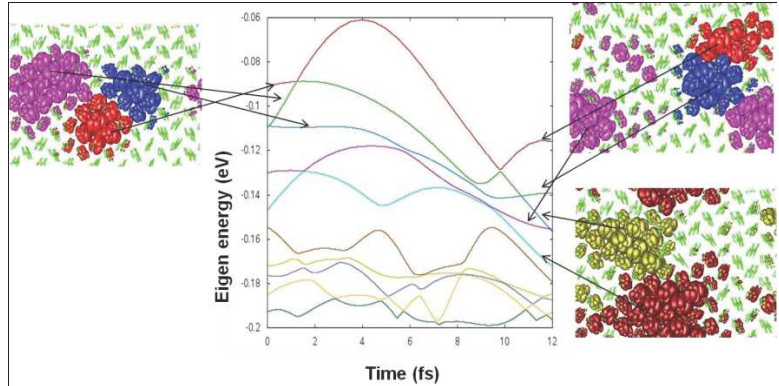

Fig.4, the adiabatic state eigen energies (center) and the eigen state isosurface plots (two sides). Each colored curve in the center panel represents one $\varepsilon_i$ for one "i". The isosurface contains 98% of the charge density in $|\varphi_i(t)|^2$.

The adiabatic eigen states and eigen energies $\{\varphi_i(t),\varepsilon_i(t)\}$ are calculated for 825 fs. The calculation is carried out on the Jaguarpf machine at national center of computational science (NCCS) in Oak Ridge National Lab. The main task is to calculate $\{\varphi_i(t),\varepsilon_i(t)\}$ from Eq.(3) with an interval of 0.5 fs (thus there are 1651 snapshots for the 825 fs). In our case, we have the benefit that all the $\{R(t)\}$ are known before the calculation of $\{\varphi_i(t),\varepsilon_i(t)\}$ for this 825 fs. Thus, the calculations for different snapshots can be carried out in a parallel fashion. Typically, we carried out 22 snapshot groups at the same time. We use 2352 computer cores for each snapshot group. Thus, in total, we have used 51,744 computer cores. In the OFM calculation, the 2352 cores for each snapshot groups are further divided into 294 groups, each group with 8 cores. In this way, one computer core group (with 8 cores) will calculate one fragment in OFM. Each snapshot group will typically have 25 (due to memory limitation) snapshots consecutively in time. In this way, one snapshot's calculation can start with the fragment wave functions of the previous snapshot, thus saving time for the iterative convergence. One such calculation takes about 1 hour of wall clock time and we can get 550 snapshots (275fs). Thus the total calculation takes about 3 hours. A plane wave kinetic energy cut off of 50 Ryd is used in the calculation, norm conservation nonlocal pseudopotentials are used, and the x, y, z real space grid points of the 7x7 supercell are 336, 252, 224 respectively.

After all the $\{\varphi_i(t),\varepsilon_i(t)\}$ are obtained, the $\psi(t)$ is integrated starting from the adiabatic VBM state at t=0. This integration part doesn't take much time. The average energy of $\psi(t)$ is shown in Fig.5(b) as the pink line. As we can see from Fig.5(b), despite the fact that at t=0, the $\psi(0)$ equals $\varphi_1(0)$, the $\psi(t)$ quickly evolves into some kind of quasi-steady state in a very short time, around 15 fs. Unfortunately, this quasi-steady state is rather unphysical. It is about 0.2 eV below the VBM, and seems to be in the middle of the 50 $\varphi_i(t)$ states chosen to be included in Eq.(4). Thus, this behavior is an artifact. This artifact is a consequence of the Ehrenfest approximation [43]. Note that so far, we have only followed the Eqs.(1) and (2), without introducing any energy surface hopping. This is an Ehrenfest dynamics. In such a dynamics, in Eq.(5), the transition from state i to state k is treated the same as the transition from state k to state i regardless of the low/high order of their eigen energies ($\varepsilon_i$, $\varepsilon_k$). As a result, the electron state $\psi$ will quickly reach an equilibrium among all the adiabatic states $\{\varphi_i\}$ in Eq.(4). This process and its transition rates between the adiabatic states violate the detailed balance (which states that the ratio of the averaged transition rate from i to k and the transition rate from k to i should be equal to $\exp(-(\varepsilon_i-\varepsilon_k)/kT)$). This problem of Ehrenfest dynamics has been discussed in Ref.[25], although the solution of this problem given in that work is different from the one we will provide below.

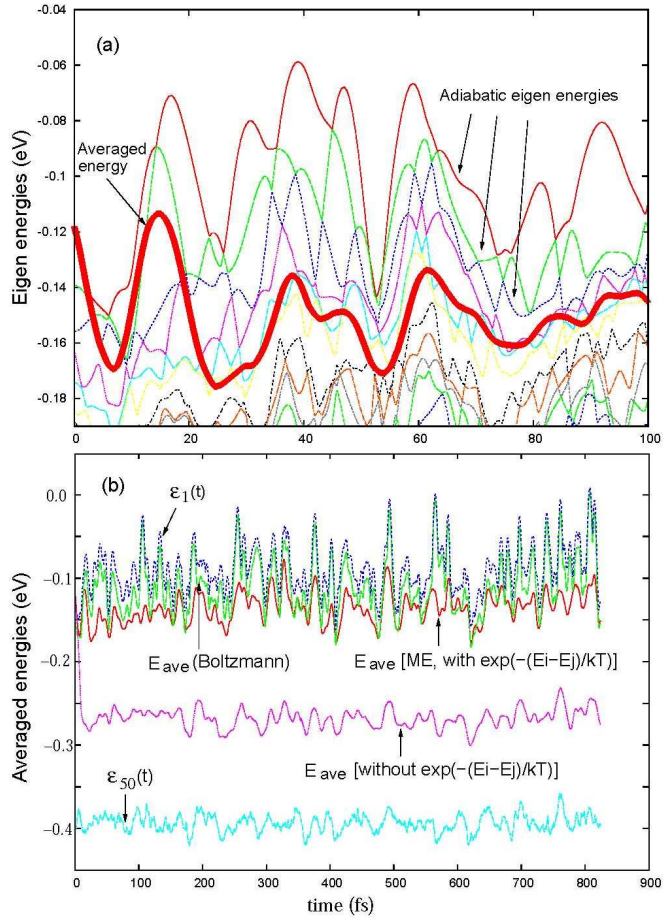

Fig.5, the adiabatic eigen energies and the averaged $\psi(t)$ energy. (a) is the enlarged view for only 100 fs, the thin lines are adiabatic eigen energies, and the thick red line is the averaged $\psi(t)$ of the modified Ehrenfest (ME) method using Eq.(8). (b) is for the full simulation plotting out only $\varepsilon_1(t)$ and $\varepsilon_{50}(t)$. The red line is the same as the thick red line in (a), while the pink line is original Ehrenfest method without the Boltzmann factor in Eq.(8), and green line is the average energy calculated using Boltzmann distribution from all the adiabatic eigen energies $\{\varepsilon_i(t)\}$.

Note that, our approximation which detaches the nuclei dynamic of Eq.(2) from the electronic movement of Eq.(1) (as a result, the total energy is not conserved) is not the source of this problem. To test this, we have taken a simple 1D model Hamiltonian from Ref.[30] which includes both the electronic and nuclei dynamics. We have performed an Ehrenfest dynamics of this system which conserves the total energy. The result also shows similar behavior as in the dynamics shown in Fig.5(b). This problem is a consequence of the classical treatment of the nuclei dynamics using Newton's law. In a quantum mechanical treatment of the nuclei movement, even at zero temperature (no atomic movement in the classical treatment), there is a zero phonon mode, which can still induce transition from i to k as long as $\varepsilon_i < \varepsilon_k$ (for hole state energy, this means the hole loses energy in the transition), but the transition from k to i cannot happen. In the quantum mechanical picture, it is this difference which maintains the detailed balance between the transition rate from i to k, and the transition rate from k to i (e.g., in a simple Fermi golden rule formalism of a single phonon emission/absorption transition process, the i to k transition rate has a prefactor $(n_v+1)$, the k to i transition rate has a prefactor $n_v$, here $n_v$ is the Bose-Einstein distribution of the phonon mode $v$ with an energy $\varepsilon_k-\varepsilon_i$ ). All these are missing in the classical treatment of the nuclei movement.

In Tully's algorithm [44], this detailed balance is restored by checking the possibility of each energy surface hopping. In Tully's algorithm and the related picture, although the $\psi(t)$ is calculated, the actual electronic system always resides on one adiabatic state $\varphi_i(t)$ (thus on an adiabatic energy surface) at a given time t and its trajectory keeps hopping (switching) among the different i's (hence the name: energy surface hopping). Thus the electronic property of the system at time t should be calculated based on $\varphi_i(t)$ (although the index i can suddenly change with time), but not on $\psi(t)$. The surface hopping from i to k is sudden (instantaneous), but the $\psi(t)$ will not change after the hopping, only the nuclei movement will be rescaled, and follow a different energy surface (adiabatic state). In a way, the $\psi(t)$ is just an auxiliary function to help us to determine how does the system hop from one adiabatic state i to another adiabatic state k. In the mean time, between two hoppings when the system resides on an adiabatic state i, the nuclei movement follows the energy surface of this adiabatic state i (just as in a Born-Oppenheimer approximation, although here i might not be the ground state), thus the total energy is also conserved. Note, this is a very different picture than the picture in Ehrenfest dynamic, where the real electronic system is described by $\psi(t)$, not just on one adiabatic state, and the electronic properties should be calculated based on $\psi(t)$.

In Tully's algorithm, when a state hopping from i to k is proposed to happen, a related transition degree of freedom in the nuclei system is calculated, then the kinetic energy on this degree of freedom is checked. If $\varepsilon_i < \varepsilon_k$, then the hopping will always be allowed, and the velocity on the transition degree of freedom will be rescaled so its kinetic energy will increase by $\varepsilon_k-\varepsilon_i$ after the hopping. If $\varepsilon_i > \varepsilon_k$, and the related kinetic energy is smaller than $\varepsilon_i-\varepsilon_k$, then the hopping will be forbidden (not happen). If the kinetic energy is larger than $\varepsilon_i-\varepsilon_k$, then the hopping will happen, but the velocity will be rescaled so its kinetic energy will be reduced by $\varepsilon_i-\varepsilon_k$. This procedure ensures that the total energy is conserved during the transition, and it treats the transition from i to k differently from the transition from k to i depending on their relative energies. This different treatment restores the detailed balance.

In our detached treatment between the electron and nuclei movements, although the total energy conservation as the sum of the electron and nuclei systems is no longer held, we can still use the nuclei kinetic energy on the transition degree of freedom to determine whether an energy surface hopping is allowed or forbidden following Tully's algorithm (we might choose not to rescale the nuclei kinetic energy after the transition). Since the kinetic energy on any degree of freedom follows the Boltzmann distribution, thus statistically, it is equivalent to say when $\varepsilon_i < \varepsilon_k$, the transition is always allowed, and when $\varepsilon_i > \varepsilon_k$, the transition only has a probability of $\exp(-(\varepsilon_i-\varepsilon_k)/kT)$ for it to happen. This procedure has been used by Prezhdo et.al [19] to study the carrier cooling in a quantum dot. Here, we will follow the same procedure. Note that, in this case, all the effects of surface hopping can be described by a master equation on the populations $P_i(t)$ on each $\varphi_i(t)$ state, while the transition rate from $P_i(t)$ to $P_k(t)$ is described by Tully's formula [44]: $R'_{i,k}(t)=Re[C_k(t)C_i^*(t)V_{ik}(t)]/|C_i(t)|^2$ with an additional prefactor 1 for $\varepsilon_i < \varepsilon_k$ or $\exp(-(\varepsilon_i-\varepsilon_k)/kT)$ for $\varepsilon_i > \varepsilon_k$.

Before showing the result of Tully's algorithm, we would first like to introduce an alternative approach. In this approach, we follow the physical picture of Ehrenfest approximation, e.g., the electronic system is described by $\psi(t)$, and there is no energy surface hopping. However, in order to describe the effect of the quantum mechanical zero phonon mode, we will introduce a factor in Eq.(5), thus we have:

$$\dot{C}_i(t) = -i\varepsilon_i(t)C_i(t) - \sum_k C_k(t)V_{ik}(t)f(\varepsilon_i - \varepsilon_k, t) \tag{8}$$

Here, if $R_{i,k} \equiv Re[C_k(t)C_i^*(t)V_{ik}(t)] > 0$ (the weight transition is from i to k), then $f(x,t)=1$ for $x<0$ (energy of state i is larger than the energy of state k) and $f(x,t)=\exp(-x/kT)$ for $x>0$ (energy of state i is smaller than the energy of state k). Similarly, if $R_{i,k}<0$ (the weight transition is from k to i), then $f(x,t)=\exp(x/kT)$ if $x<0$ and $f(x,t)=1$ if $x>0$. We will call the algorithm of Eq.(8) as the modified Ehrenfest (ME) algorithm. Note the $\psi(t)$ in this algorithm will be different from the $\psi(t)$ in Tully's algorithm which follows Eq.(5). One might worry that the change of $f(x,t)$ might interrupt the coherence of the wave function evolution. Such coherence of wave function change (e.g., the coherent accumulation of the wave function weight from one adiabatic state to another adiabatic state) is important, for example to describe a single phonon absorption/emission induced transition in the weak coupling case. More in-depth analysis can show that such concerns are not warranted. For example, in a state energy crossing case, where the weight transition from one state to another can be large and rapid, the $|\varepsilon_i-\varepsilon_k|$ usually is very small, thus the effects of the factor $f(x,t)$ is small. In the case of one phonon absorption (or emission) induced transition, usually the sign of $Re[C_k(t)C_i^*(t)V_{ik}(t)]$ does not change in the whole transition period (where coherent weight accumulation happens), thus the $f(x,t)$ also does not change during this period, so the coherence of the wave function change will not be interrupted. We do note that, there are other intrinsic shortcomings of the Ehrenfest algorithm [43,44,45]. Most notably the spreading of $\psi(t)$ overall several $\varphi_i(t)$ which are physically separated. That will affect the amplitudes of the polaronic atomic relaxations. In our current work, we have ignored such polaronic effect from the beginning, thus the corresponding shortcoming is not an issue here.

The result of this modified Ehrenfest (ME) algorithm is shown in Fig.5(a) and (b) as the red lines. We see that, this time, the average energy of $\psi(t)$ is much closer to the VBM energy as expected. In Fig.6, we also show the comparison between the ME result and the Tully's algorithm result [note, in Tully's algorithm, the ψ(t) is still described by Eq.(5) and the average energy of ψ(t) is the pink line in Fig.5(b), but the populations on the adiabatic states have a much higher average energy]. From the plot, we see that although ME algorithm and Tully's algorithm give different results, they are close to each other, and behave qualitatively in a similar way.

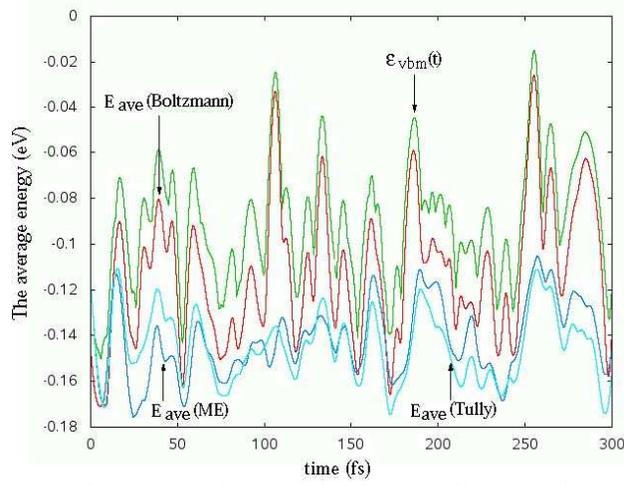

Fig.6, the comparison between the ME algorithm result and the Tully's algorithm result.

At this time, it is interesting to introduce the Boltzmann average energy. This is the average energy at time t according to a Boltzmann distribution of the adiabatic states $\varepsilon_i$: $\varepsilon_{ave} = \sum_i \varepsilon_i \exp(\varepsilon_i/kT) / \sum_i \exp(\varepsilon_i/kT)$. This energy is shown in Fig.5(b) and Fig.6. We can see that the Boltzmann average energy is higher than the two NAMD results, and it is much closer to the VBM. This is an important conclusion because in many phenomenological treatments, the Boltzmann distribution among the adiabatic electron states is always assumed. Here the reason for the NAMD results (either the ME algorithm or Tully algorithm) to fail to reach the Boltzmann distribution is due to the rapid eigen energy changes of the adiabatic states, especially for the band edge VBM state. The system does not have enough time to respond to this change, hence to reach the equilibrium, thus it is always in a nonequilibrium situation.

To further illustrate the difference between the ME and Tully algorithms, we have shown in Fig.7 the population $P_i(t)$ on the adiabatic state i in the Tully's algorithm, and the $|C_i(t)|^2$ of Eq.(4) in the ME algorithm. We see that, the $P_i(t)$ and $|C_i(t)|^2$ can change suddenly, but that is due to the state crossing, where the indexes of the two crossing states exchanged. At the beginning, when $|C_i(t)|^2$ and $P_i(t)$ started from the same occupation (e.g., VBM), the first few states in ME and Tully's algorithm look similar. But when their amplitudes become small, their behaviors are quite different. This means the $P_i(t)$ in Tully's algorithm and $|C_i(t)|^2$ in ME algorithm can be very different when both of them are very small. Note, both $P_i(t)$ in Tully algorithm and $|C_i(t)|^2$ in ME algorithm are mainly distributed among 4-7 states near the top of the valence band. Given these differences, at this stage, it is still difficult to judge which algorithm is better, making more physical sense.

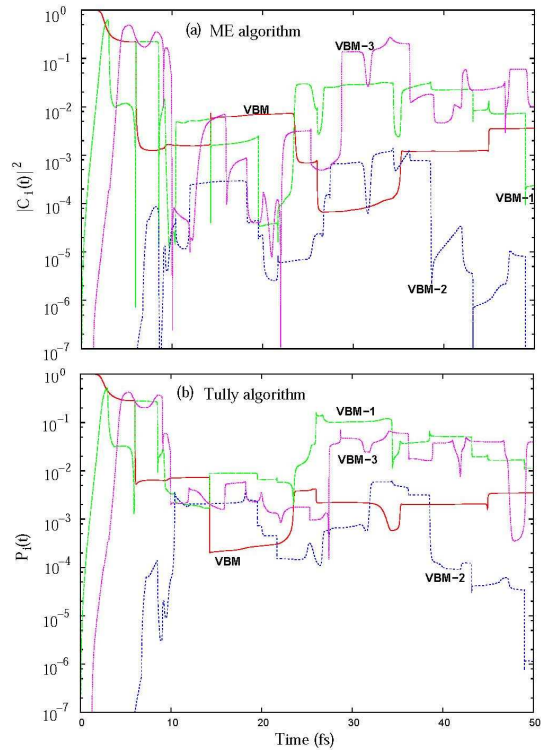

Fig.7, $|C_i(t)|^2$ in the ME algorithm (a), and $P_i(t)$ in Tully's algorithm (b). A sudden change in these quantities usually means a state crossing (as a result, the identity of the states exchanges).

## 4. The simulation results and the carrier transport mechanisms

The main purpose of our study is to reveal the underlying mechanism of the carrier transport in the monolayer thin film of 5TBA. Due to its molecular crystal structure, this system can be used as an example for other organic molecular crystals. We do have to keep in mind that the relative strength between the inter-molecular coupling, the molecule reorganization energy, and thermo fluctuation effect can divide the organic molecular crystals into different regimes. Our case might belong to relatively large thermo fluctuation and inter-molecular coupling. Prior to the calculation, one can hypothesize different possible mechanisms for the carrier transport: (i) band structure transport by extended bulk Bloch states [3]; (ii) polaron hopping, where the polaron localization is induced by atomic relaxation due to the existence of the hole state [6]. Furthermore, many of the literature for such molecular crystals assume the polaron is one hole at one molecule (in our case one 5TBA oligomer) [46]; (iii) localized state drifting, where the localization is induced by molecule thermo fluctuation, and the same fluctuation can cause the localized states to change their positions with time. Thus the carrier mobility is produced by the state shifting (while the hole is residing on the same state without any state transition); (iv) localized state transitions by absorption/emission of a single phonon as described by a simple Fermi golden rule [5]; (v) localized state energy crossing, while they cross each other in energy, the carrier residing on one state can jump to another state.

We see many band structure calculations for organic periodic structures. The underline assumption could be that the band structure and the effective mass can be relevant to the carrier transport through a bulk transport picture. Our calculation shows that at room temperature, the adiabatic states are localized among 10-20 5TBA oligomers as shown in Fig.4. Thus the wave function is not extended, and the hypothesis (i) is invalid. Another common picture is that the wave function will localize in a single unit (e.g., a molecule, or the oligomer in our case) of the structure and form a polaron. Then it will hop from one unit to another. In such a polaron picture, the localization is induced by atomic relaxation due to strong electron-phonon coupling (not by thermo fluctuation), thus it will form even at zero temperature. We have performed DFT/LDA calculations with the 2x2 supercell (which contains 8 5TBA oligomers) containing one hole at zero temperature. Despite initial distortion to the atomic positions and initial hole wave function localizations, after SCF iterations and atomic relaxation, the hole wave function is extended uniformly in the 2x2 supercell and there is no localization. This means that at zero temperature, there is no polaron, at least no polaron with wave functions localizations smaller than 8 5TBA oligomers. If there is a polaron, the size of the polaron must be larger than 8 5TBA oligomers, hence larger than the thermo fluctuation induced localization as shown in Fig.4. Thus the hypothesis (ii) does not hold, and the main reason for localization is the thermo fluctuation (dynamic disorder), not the hole induced atomic relaxation and the reorganization energy. From the 2x2 supercell DFT calculation, we get a reorganization energy of 20 meV (the reorganization energy is the energy drop of the system with a hole from the initial neutral system (without hole) relaxed atomic positions to the final relaxed atomic positions with the hole). Since the thermo fluctuation induced wave function localizations shown in Fig.4 are similar to the 2x2 supercell size, we expect that the atomic relaxation energies of these localized states should also be around 20 meV. Such atomic relaxation energy, e.g. the electron wave function feedback to nuclei movement, is ignored in our current detached treatment, thus we don't have this 20 meV relaxation energy. Nevertheless, this energy is much smaller than the ~100 meV energy fluctuations of VBM shown in Figs.5, 6, which is induced by thermo fluctuation. Once again, this shows that the polaron effect is much weaker than the thermo fluctuation effect in our system. Nevertheless the inclusion of such "polaron" relaxation effect in the future might modify the results quantitatively to some degree, although we do not expect it to change the qualitative picture. In the following, we will address the points (iii), (iv) and (v).

Figure 8 shows the diffusion distance square as a function of time t. Note, although this is done with t=0

not in the steady state (e.g., $\psi(t=0)=\varphi_{VBM}(t=0)$), recalculating this diffusion distance starting from a steady state (e.g., after 10 fs) snapshot doesn't change the result very much, probably due to the fast realization of the steady state near t=0. Note, the 7x7 supercell is still relatively small. To calculate the diffusion distance, we have repeated the system a few times in its x and y directions, and treat the image states of an adiabatic state $\varphi_i(t)$ as different states (e.g., with different coefficient $C_i(t)$ in Eq.(4) and (7), and different $P_i(t)$ in Tully's algorithm). This allows us to have larger diffusion distance than the box size of the 7x7 supercell. Nevertheless, there are some limitations of this technique, as a result the diffusion distance is saturated both for the ME algorithm and the Tully's algorithm. Thus we should only judge the carrier mobility from the time period before the saturation.

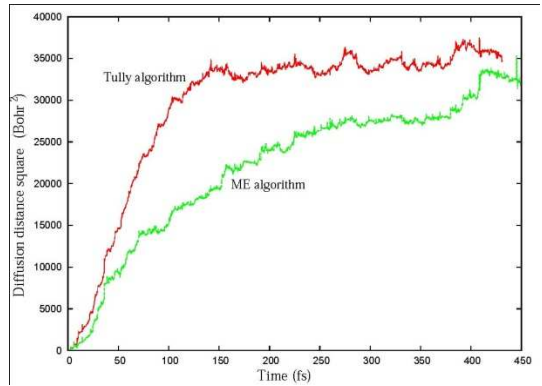

Fig.8, the diffusion distance square as a function of the simulated time t.

From Fig.8, we see that initially, the ME and Tully's algorithms get similar results. However, after some time (50 fs), the Tully's algorithm yields a bigger diffusion distance. It is interesting to speculate whether the slowdown in the ME algorithm is related to the weak localization phenomenon in a disordered system, where coherently constructive back scattering can slow down the carrier diffusion [47]. Further analysis for this point will be done in the future. From the slop of the line in Fig.7 (taken from the Tully algorithm result), if we use a 2D diffusion formula of $d^2=4Dt$, and $\mu=eD/kT$, we get a hole mobility $\mu$ as 44cm$^2$/Vs. This is a little bit large considered that most organic crystals have mobilities between 1-10cm$^2$/Vs. But we have to keep in mind that this is a 2D system. We are still waiting for experimental confirmation for this result. Another possibility is that, the polaronic atom relaxation effect discussed above might be able to induce some additional self trapping for the thermo fluctuation induced localized states. Such deeper trapping can reduce the mobility. Further investigation is needed to know how large is such effect. We have also investigated the x and y direction diffusions, and found that the x direction is diffusing faster than the y direction [see Fig.1S].

Now, to answer whether the diffusion is caused by state position drifting, we have plotted the center of the mass position of the first adiabatic state (VBM) in Fig.9 as a function of the time. We see that when there is no state crossing, the position of the state does not change, despite of the fact, the energy of the state can change by as much as 0.1 eV during the same period. The same is confirmed by looking at the states in Fig.4 (e.g., the second state at t=0, and the third state at t=12 fs in Fig.4). In Fig.9, when the VBM and VBM-1 states cross, the identities of the states switch, which can cause a sudden change of the VBM position in the plot. Nevertheless, from the flat plateau in the state position between the state crosses, we can conclude that the state position drifting effect proposed in hypothesis (iii) should be small.

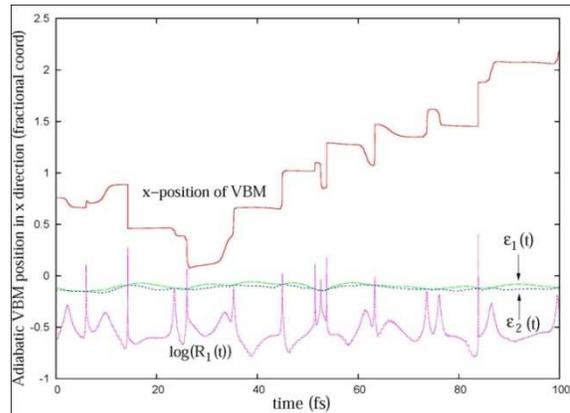

Fig.9, the center of mass position of the VBM state. Also shown are the eigen energies of the first two adiabatic states, and the total transition rate $R_1(t)$ (defined as $R_i \equiv \sum_k |R_{i,k}|/|C_i|^2$ ) from the VBM state. When there is a peak in the transition rate $R_1(t)$, there is a state crossing between the VBM and VBM-1 states. Also, at such crossing, the VBM state position suddenly changes, which is due to the change of identity between the VBM and VBM-1 state (the highest adiabatic state is called VBM), rather than a genuine shift of the state position.

Next, we have investigated the state transition by absorbing or emitting a single phonon. Such transition is used to explain the carrier mobility in a disordered polymer system [5]. But here, the adiabatic state localizations are caused by thermo fluctuations, not by the disorder tangling/arrangement of polymer chains. Thus the state here can change more rapidly than in the disordered polymer system. In the Tully algorithm (similar things can be done for the ME algorithm), we can turn off any energy surface hopping between states i and k, when $|\varepsilon_i-\varepsilon_k|>5$meV. Since most phonon modes have a phonon energy larger than 5 meV, this will effectively turn off all the single phonon absorption/emission effects. The resulting diffusion distance square is very similar as the original result as shown in Fig.10. This means the effect of single phonon absorption/emission transition as described in hypothesis (iv) is rather small, and can be ignored. Thus, finally, we can conclude that all the transitions happen when two states cross each other in their energies. When that happens, there is a fast transition of the wave function weight from one state to another state. That causes the position of the carrier to suddenly jump, and this jump (as described in hypothesis (v)) is the real cause for carrier mobility in our system.

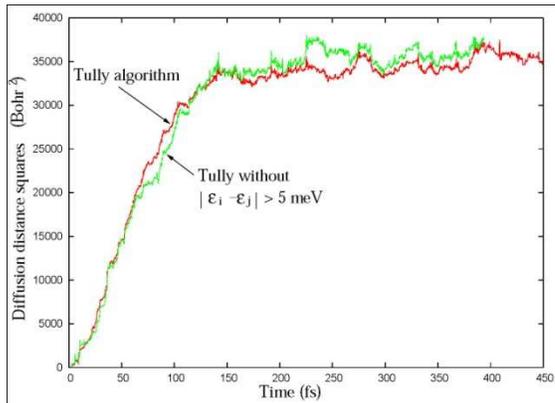

Fig.10, the diffusion distance with and without the $|\varepsilon_i-\varepsilon_j|>5$meV transitions.

It is interesting to comment on the difference and similarity of the state crossing observed in our simulation and the state crossing in Marcus theory [32]. In both pictures, the state crossing plays an important role, and the transition happens when there is a state crossing. But in Marcus theory, the atomic relaxation related reorganization energy plays an important role. The localization is either formed by a geometric confinement (e.g., within a molecule, or inside a quantum dot), or it is due to this reorganization (e.g., in a polaron). The reorganization energy is also the parameter to gauge the energy fluctuation which causes the state crossing. On the other hand, in our current problem, the localization of the state and the state energy fluctuation are caused by dynamic disorder induce by thermo fluctuation (not by reorganization energy). The fact that we get such state crossing while ignoring the hole induced atomic relaxation indicates that there are fundamental differences between these two pictures. We also expect very different temperature dependence on the mobility (in our case, the state localization, not just the state energy fluctuation, also depends on the temperature).

It is also interesting to discuss the relationship between our picture and the dynamic disorder picture proposed by Troisi, et.al [19,30]. In their simulations, a model Hamiltonian is used which includes the fluctuation of the inter-molecule coupling. They have used pure Ehrenfest dynamics to monitor the diffusion of the electron wave function. Both in their picture and our picture, the carrier mobility is driven by this thermo fluctuation (dynamic disorder). However, in our simulation, we have included the surface hopping in Tully's algorithm, or the additional Boltzmann factor in the ME. All these rely on the representation of the wave function $\psi(t)$ in terms of adiabatic states $\{\varphi_i(t)\}$. As a result, we can introduce the concept of state crossing, and point out that the state crossing is the main underlying mechanism for the diffusion. This is absent from a pure Ehrenfest dynamics treatment. We have also used a realistic DFT Hamiltonian, which includes both intra and inter molecular fluctuations, and point out the described phenomenon happens in real systems.

## 5. Conclusion

In conclusion, we have presented several techniques to carry out a large scale simulation for non-adiabatic molecular dynamics (NAMD). These techniques and supercomputer facilities allow us to do NAMD in ab initio level for a 4802 atom system for about 1 ps in a few hours. More specifically, (i) a linear approximation of the Hamiltonian dependence on time t is introduced, which increases the time step of the computationally most expensive part, from $10^{-3}$ fs to 0.5 fs, thus reduces the computational cost by hundreds of times. Similar techniques can probably be used in simulations like real time TDDFT; (ii) a modified Ehrenfest dynamics is proposed, which restores the detail balance but without the energy surface hopping. It can be used as an alternative approach to the Tully's algorithm; (iii) the nuclei movement and the electronic movement are detached, and classical force field is used for the nuclei movement. The charge patching method is used to construct the Hamiltonian, while the overlapping fragment method is used to diagonalize the Hamiltonian matrix; (iv) the whole approach can be carried out using massively parallel computers, and can be scaled to 50,000 cores. As a result, the whole calculation takes only a few hours.

We have used this approach to study the carrier transport of a 5TBA monolayer thin film. Such thin film has been synthesized experimentally, and its transport property is planned to be measured experimentally. Our main task is to reveal the underlying mechanism of hole transport in such a system. This system demonstrates a herringbone 2D crystal structure, it thus can be used as one prototype for a class of organic molecular crystals. Through our calculations, we found a new mechanism for carrier transport. The carrier transport is mostly induced by state crossing between localized states. The localizations of the states are caused by the dynamic disorder induced by thermo fluctuation of the molecules, rather than by polaronic atomic relaxation induced by the existence of the hole. Thus the state crossing mechanism is very different from the ones described by Marcus theory. The state localization size is about 10-20 5TBA oligomers, instead of one 5TBA oligomer. In our simulation, we also found that the occupations of the adiabatic states are often in a non-equilibrium situation. The average energy calculated by the Boltzmann distribution using the adiabatic state eigen energies is much closer to the VBM energy than the NAMD simulated result. The reason for this is the rapid fluctuations of the adiabatic state eigen energies induced by thermo fluctuation. The electronic state weight redistribution cannot respond to this eigen energy change fast enough to reach the equilibrium. Finally, our calculated hole mobility is 44 cm$^2$/Vs, which is awaiting for experimental confirmation.

ACKNOWLEDGMENT

We like to thank Dr. Martin, Hendriksen, and Salmeron for stimulating discussion. This work was supported by SC/BES/MSED of the U.S. Department of Energy under the Contract No. DE-AC02-05CH11231. This research used resources of the Oak Ridge Leadership Computing Facility, located in the National Center for Computational Sciences at Oak Ridge National Laboratory, which is supported by the Office of Science of the Department of Energy under Contract DE-AC05-00OR22725. The computer time was allocated by the Department of Energy's Innovative and Novel Computational Impact on Theory and Experiment (INCITE) program. Vukmirovic is supported by European Community FP7 Marie Curie Career Integration Grant (ELECTROMAT), Serbian Ministry of Science (ON171017) and FP7 projects PRACE-1IP, PRACE-2IP, HP-SEE and EGI-InSPIRE

# Supplementary information for: Non-adiabatic Molecular Dynamics Simulation for Carrier Transport in a Molecular Monolayer

*Junfeng Ren*[(1,2)], *Nenad Vukmirović*[(3)], *Lin-Wang Wang*[(1)*]

(1) Material Science Division, Lawrence Berkeley National Laboratory, One Cyclotron Road, Mail Stop 66, Berkeley, CA 94720

(2) College of Physics and Electronics, Shandong Normal University, Jinan 250014, China

(3) Scientific Computing Laboratory, Institute of Physics Belgrade, University of Belgrade, Pregrevica 118, 11080 Belgrade, Serbia

(*) Email: lwwang@lbl.gov

Fig.1s shows the diffusion distance squares in x and y directions defined in Fig.1. These are calculated by the modified Ehrenfest dynamics. The results from Tully's algorithm are similar. Note that the ratio of the diffusion distance squares (hence the mobility) in the x and y directions is the opposite as reported in Ref.[1]. This is because in Ref.[1], the oligomer is TD5TBA (tetradecyl pentathiophene butyric acid), while in the current work, the oligomer is 5TBA (pentathiophene butyric acid) without the top alkane chain. Due to this difference, the packing lattice constants are different for these two oligomers. In 5TBA, the packing in x direction is closer, resulted in stronger electron coupling and higher mobility in this direction.

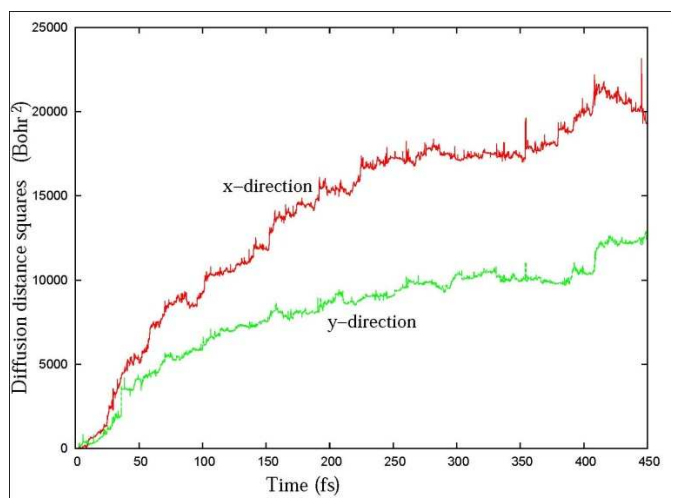

Fig.1S, the diffusion distance squares in the x and y directions calculated by the modified Ehrenfest algorithm.